\documentclass[sigconf]{acmart}
\AtBeginDocument{%
  \providecommand\BibTeX{{%
    \normalfont B\kern-0.5em{\scshape i\kern-0.25em b}\kern-0.8em\TeX}}}

\usepackage{multirow}
\usepackage{float} 
\usepackage{amsmath}
\usepackage[ruled,linesnumbered]{algorithm2e}
\usepackage{graphicx}
\usepackage{subfigure}
\usepackage{epstopdf}
\usepackage{tablefootnote}
\usepackage[flushleft]{threeparttable}
\usepackage{diagbox}
\usepackage{booktabs}
\usepackage{siunitx}
\usepackage{footmisc}
\usepackage{footnote}
\usepackage{enumitem}
\usepackage[normalem]{ulem}
\useunder{\uline}{\ul}{}

\copyrightyear{2023}
\acmYear{2023}
\setcopyright{rightsretained}\acmConference[COLIEE 2023]{COLIEE 2023 workshop: Competition on Legal Information Extraction/Entailment}{June 19, 2023}{Braga, Portugal}
\acmBooktitle{Proceedings of COLIEE 2023 workshop, June 19, 2023, Braga, Portugal}
\acmPrice{}
\acmISBN{}
\acmDOI{}

\begin{document}

\title{THUIR@COLIEE 2023: Incorporating Structural Knowledge into Pre-trained Language Models for Legal Case Retrieval}

\author{Haitao Li}
\affiliation{DCST, Tsinghua University}
\affiliation{Quan Cheng Laboratory}
\affiliation{Beijing 100084, China}
\email{liht22@mails.tsinghua.edu.cn}

\author{Weihang Su}
\affiliation{DCST, Tsinghua University}
\affiliation{Quan Cheng Laboratory}
\affiliation{Beijing 100084, China}
\email{swh22@mails.tsinghua.edu.cn}

\author{Changyue Wang}
\affiliation{DCST, Tsinghua University}
\affiliation{Quan Cheng Laboratory}
\affiliation{Beijing 100084, China}
\email{changyue20@mails.tsinghua.edu.cn}

\author{Yueyue Wu}
\affiliation{DCST, Tsinghua University}
\affiliation{Quan Cheng Laboratory}
\affiliation{Beijing 100084, China}
\email{wuyueyue@mail.tsinghua.edu.cn}

\author{Qingyao Ai}
\affiliation{DCST, Tsinghua University}
\affiliation{Quan Cheng Laboratory}
\affiliation{Beijing 100084, China}
\email{aiqy@tsinghua.edu.cn}

\author{Yiqun Liu}
\authornote{Corresponding author}
\affiliation{DCST, Tsinghua University}
\affiliation{Quan Cheng Laboratory}
\affiliation{Beijing 100084, China}
\email{yiqunliu@tsinghua.edu.cn}
\renewcommand{\shortauthors}{Haitao Li et al.}
\begin{abstract}
Legal case retrieval techniques play an essential role in modern intelligent legal systems. As an annually well-known international competition, COLIEE is aiming to achieve the state-of-the-art retrieval model for legal texts. This paper summarizes the approach of the championship team THUIR in COLIEE 2023. To be specific, we design structure-aware pre-trained language models to enhance the understanding of legal cases. Furthermore, we propose heuristic pre-processing and post-processing approaches to reduce the influence of irrelevant messages. In the end, learning-to-rank methods are employed to merge features with different dimensions. Experimental results demonstrate the superiority of our proposal. Official results show that our run has the best performance among all submissions. The implementation of our method can be found at https://github.com/CSHaitao/THUIR-COLIEE2023.
\end{abstract}



\maketitle

\section{Introduction}
In countries with case law systems, precedent is an important determinant for the decision of new given cases~\cite{locke2022case,shao2023understanding}. Therefore, it takes a substantial amount of time for legal workers to find precedents that support or contradict a new case. With the growing number of digital legal cases, it is increasingly more expensive for legal practitioners to find precedents. Recently, the growing works have raised the awareness that legal search systems will free people from the heavy manual work~\cite{shao2020bert,bench2012history,yu2022explainable,ma2021lecard,althammer2021dossier,ma2023caseencoder}.

In ad-hoc retrieval and open-domain search, contextual language models such as BERT have brought significant performance gains to the first stage of retrieval~\cite{xie2023t2ranking}. Despite their great success, applying language models to legal case retrieval is not trivial with the following main challenges.

\begin{table*}[t]
\caption{Dataset statistics of COLIEE Task 1.}
\begin{tabular}{l|cc|cc|cc}
\hline
\multirow{2}{*}{}                        & \multicolumn{2}{c|}{COLIEE 2021} & \multicolumn{2}{c|}{COLIEE 2022} & \multicolumn{2}{c}{COLIEE 2023} \\
                                         & Train           & Test           & Train           & Test           & Train           & Test          \\ \hline
\# of queries                            & 650             & 250            & 898             & 300            & 959             & 319           \\
\# of candidate case per query             & 4415            & 4415           & 3531            & 1263           & 4400            & 1335          \\
avg \# of relevant candidates/paragraphs & 5.17            & 3.6            & 4.68            & 4.21           & 4.68            & 2.69          \\ \hline
\end{tabular}
\label{satistics}
\end{table*}

Firstly, it is labor-intensive to construct high-quality annotated datasets for legal case retrieval due to the need for legal knowledge. Hence, the current dataset usually has only a few thousand training data, which may lead to over-fitting of the language model. Secondly, legal cases are usually long texts with internal writing logic. To be specific, legal cases usually contain three parts: Fact, Reasoning, and Decision. The Fact section describes the defendant's and plaintiff's arguments, evidence, and basic events. The
 Reasoning section is the analysis by the judges of the legal issues in the facts. The Decision section is the specific response of the court to all legal issues. Limited by the input length of 512 tokens, existing language models either truncate the redundant content or flatten the input of all structures, making it difficult to understand legal cases properly.

To tackle the above challenges, we propose SAILER~\cite{li2023sailer}, which stands for \textbf{S}tructure-\textbf{A}ware pre-tra\textbf{I}ned language model for \textbf{LE}gal case \textbf{R}etrieval. SAILER utilizes an encoder-decoder architecture to explicitly model the relationships between different structures and learns the legal knowledge implied in the structures through pre-training on a large number of legal cases.

To verify the effectiveness of SAILER, the THUIR team participates in the COLIEE 2023 legal case retrieval task and wins the championship. This paper elaborates on our technical solutions and demonstrates the effectiveness of incorporating structural knowledge into pre-trained language models.

The remainder of the paper is organized as follows: Section 2 introduces the
background for legal case retrieval and dense retrieval. Section 3 presents the description, datasets, and evaluation metrics of the COLIEE 2023 legal case retrieval task. In Section 4, the technical details are elaborated. After that, Section 5 introduces the experiment results. Finally, we conclude this paper in Section 6 by summarizing the major findings and discussing future work.

\section{Related Work}
\subsection{Legal Case Retrieval}
Legal case retrieval, which aims to identify relevant cases for a given query case, is a key component of intelligent legal systems. A number of deep learning methods have been applied to retrieve precedents with various techniques, such as CNN-based models~\cite{tran2019building}, BiDAF~\cite{seo2016bidirectional}, SMASH-RNN~\cite{jiang2019semantic}, etc.
Recently, researchers have attempted to achieve performance gains in legal case retrieval with transformer-based language models. For example, Shao et al.~\cite{shao2020bert} propose BERT-PLI, which divides the case into multiple paragraphs and aggregates the scores together with neural networks. Furthermore, researchers have begun to design legal-oriented pre-trained models, such as Lawformer~\cite{xiao2021lawformer} and LEGAL-BERT~\cite{chalkidis2020legal}. However, neither of them design pre-training tasks for legal case retrieval. We believe that the potential of language models for legal case retrieval has not been fully exploited.

\subsection{Dense Retrieval}
Dense retrieval is a powerful retrieval paradigm that can effectively capture contextual information~\cite{dong2022incorporating,zhan2020repbert,qu2021rocketqa,fan2022pre,chen2022axiomatically,li2023constructing}. Generally speaking, dense retrieval maps queries and documents to dense embeddings with a dual encoder. Later, the inner product is applied to measure their relevance. For better performance, researchers have designed pre-trained objectives oriented to web search, which achieve state-of-the-art effectiveness. For example, Zhan et al.~\cite{zhan2021optimizing} propose dynamic negative sampling to further improve performance. Chen et al. propose ARES~\cite{chen2022axiomatically}, which attempts to incorporate axioms into the pre-training process.

\section{Task Overview}
\subsection{Task Description}
The Competition on Legal Information Extraction/Entailment (COLIEE)  is an annual international competition whose aim is to achieve state-of-the-art methods for legal text processing. There are four tasks in COLIEE 2023, and we submit systems to task 1.

Task 1 is the legal case retrieval task, which involves identifying supporting cases for the decision of query cases from the entire corpus. Formally, given a query case $Q$ and a set of candidate cases $S$, this task is to identify all the supporting cases $S_Q^* = \{S_1, S_2, ..., S_n\}$ from a large candidate pool. The supporting cases are also named ``noticed cases". For each query, participants can return any number of supporting cases that they consider relevant.

\subsection{Data Corpus}
The data corpus for Task 1 belongs to a database of case law documents from the Federal Court of Canada provided by Compass Law. Statistics of the dataset are shown in Table \ref{satistics}. From COLIEE 2021, all queries share a large candidate case pool, which is more challenging and realistic. The COLIEE 2023 dataset contains 959 query cases against 4400 candidate cases for training and 319 query cases against 1335 candidate cases for testing. 

On further analysis, we find that the average number of relevant documents per query in the training set is 4.68 while the number of relevant documents in the test set is 2.69. Therefore, we predict the top-5 possible relevant cases to calculate the evaluation metrics during training. At testing time, we adopt heuristic post-processing to avoid the performance damage caused by the inconsistent distribution of the training and testing sets. We randomly select 187 queries as the validation set and the remaining 772 queries as the training set.

\subsection{Metrics}
For COLIEE 2023 Task 1, evaluation measures will be precision, recall, and F-measure:

\begin{equation}
\text { Precision } = \frac{\# T P}{\# T P+\# F P} 
\end{equation}
\begin{equation}
\text { Recall } = \frac{\# T P}{\# T P+\# F N} 
\end{equation}
\begin{equation}
F-\text { measure } = \frac{2 \cdot \text { Precision } \cdot \text { Recall }}{\text { Precision }+ \text { Recall }}
\end{equation}

where $\#TP$ is the number of correctly retrieved candidate cases for all query cases, $\#FP$ is the number of falsely retrieved candidate cases for all query cases, and $\#FN$ is the number of missing noticed candidate paragraphs for all query cases. It is worth noting that micro-average (evaluation measure is calculated using the results of all queries) was used rather than marco-average (evaluation measure is calculated for each query and then takes average) in the evaluation process.

\section{Method}
In this section, we present the complete solution of the COLIEE 2023 Task 1. To be specific, we first perform a simple pre-processing of the data. Then, we implement traditional retrieval methods and pre-trained language models. Furthermore, we extract multiple features for each query-candidate pair. Learning-to-rank methods are employed to aggregate these features for the score. At last, we design heuristic post-processing methods to form the final submission list.

\begin{figure}[t]
\centering
\includegraphics[width=\columnwidth]{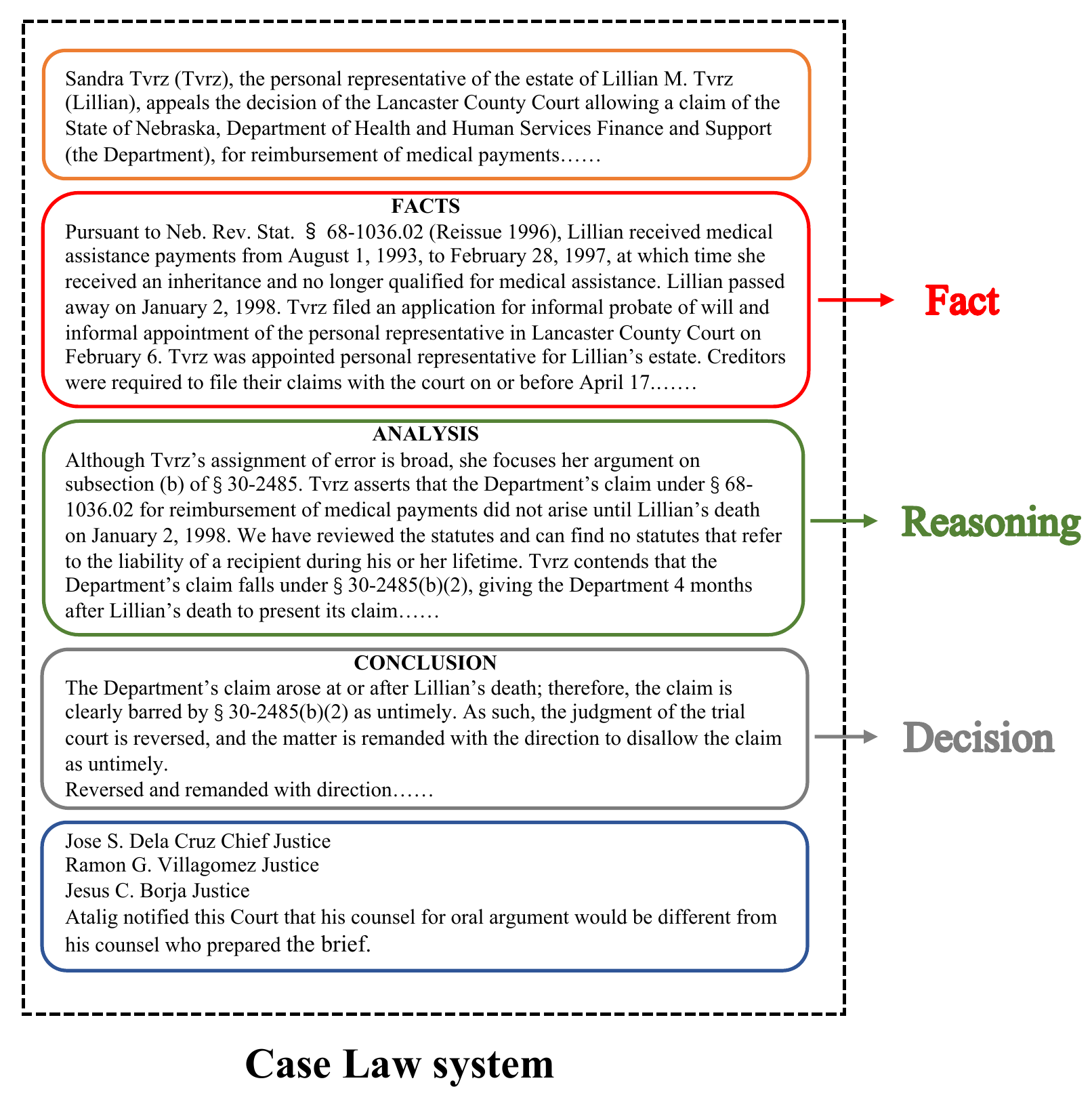}
\vspace{-0.15in}
\caption{An example of the legal case structure in the Case Law system.}
\label{example}
\end{figure}

\subsection{Pre-processing}
Before training, we perform the following pre-processing:

\subsubsection{Remove useless information}
 Firstly, we directly remove the content before character ``[1]", which is usually procedural information for that legal case, such as time, court, etc. Then, we remove the placeholders, such as ``FRAGMENT\_SUPPRESSED" etc. When calculating the similarity, these placeholders are considered as noise. Furthermore, we note that some legal cases contain French text and Langdetect is employed to remove all French paragraphs. For a few documents with a high percentage of French text, we translate them into English to retain the main information.

\subsubsection{Summary extraction}
A part of the case has the subheading of ``Summary". The summary section usually contains the important content of cases. Therefore, we extract the summary by regular matching and concatenate it at the beginning of the processed text.

\subsubsection{Reference sentence extraction}
Inspired by ~\cite{ma2021retrieving}, we are aware that placeholders such as ``FRAGMENT\_SUPPRESSED", ``REFERENCE\_SUPPRESSED",  ``CITATION\_SUPPRESSED", are citations or references from other noticed cases. These sentences are directly relevant to the supporting cases. Therefore, for all queries, we keep only the sentences with placeholders to further improve performance. Noticeably, for the candidate cases, we retain the full content.

\begin{table*}[t]
\caption{Features that we used for learning to rank. The placeholder contains ``FRAGMENT\_SUPPRESSED", ``REFERENCE\_SUPPRESSED",  ``CITATION\_SUPPRESSED".}
\begin{tabular}{cll}
\hline
\multicolumn{1}{l}{Feature ID} & Feature Name      & Description                                  \\ \hline
1                              & query\_length     & Length of the query                          \\
2                              & candidate\_length & Length of the candidate paragraph            \\
3                              & query\_ref\_num              & Number of placeholders in the query case           \\
4                              & doc\_ref\_num              & Number of placeholders in the candidate case             \\

5                              & BM25              & Query-candidate scores with BM25 (k\_1 = 3.0 , b = 1.0)             \\

6                              & QLD               & Query-candidate scores with QLD              \\
7                              & TF-IDF               & Query-candidate scores with TF-IDF              \\
8                              & SAILER        & Inner product of query and candidate vectors generated by SAILER       \\
 \hline
\end{tabular}
\label{feauture}
\end{table*}

\begin{figure}[t]
\centering
\includegraphics[width=\columnwidth]{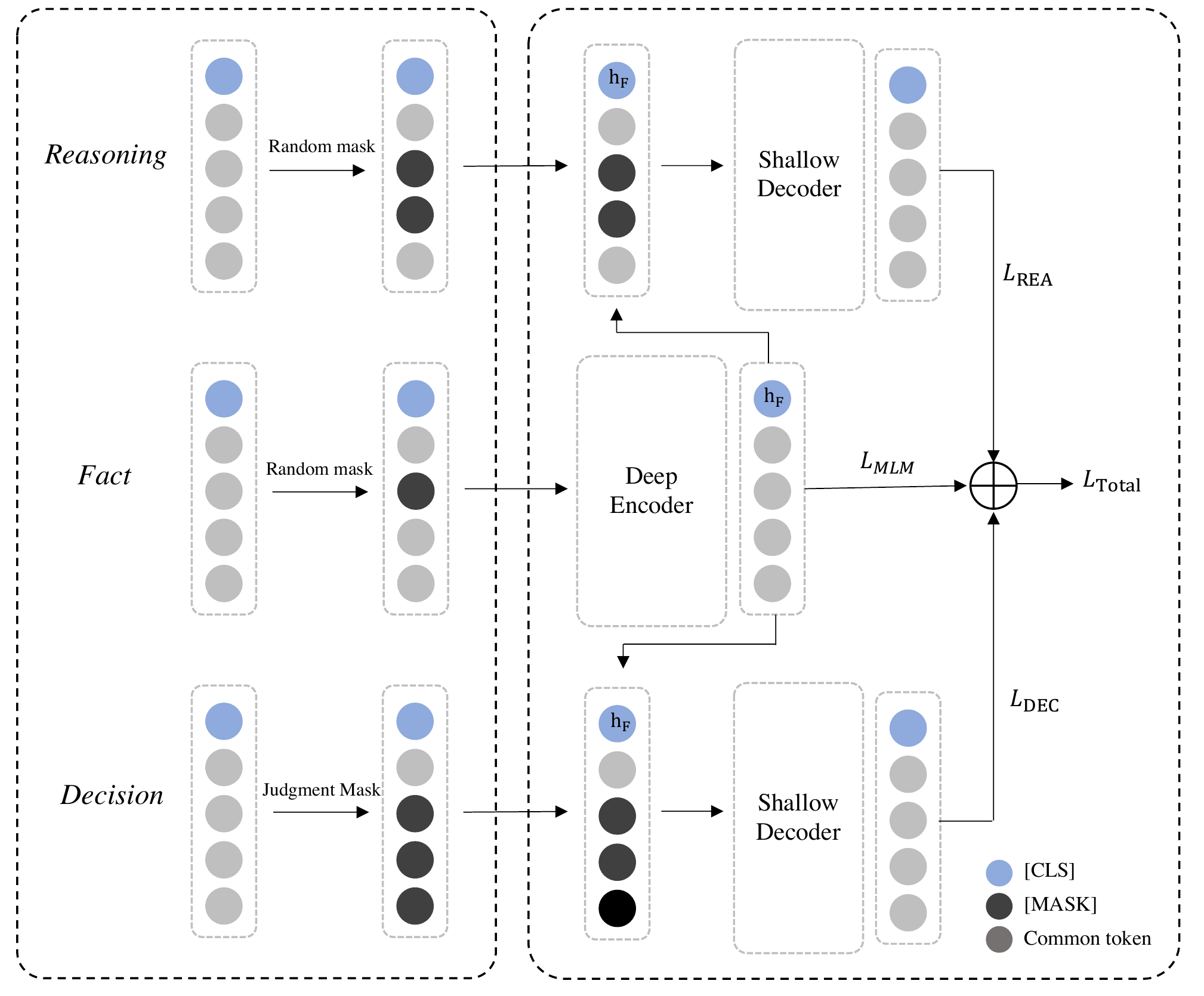}
\caption{The model design for SAILER, which consists of a deep encoder and two shallow decoders. The Reasoning and Decision section are aggressively masked, joined with the Fact embedding to reconstruct the key legal elements and the judgment results. }
\label{model}
\end{figure}

\subsection{Traditional Lexical Matching Models}
According to previous findings~\cite{ma2021retrieving,althammer2021dossier,rabelo2020coliee,rabelo2022overview}, the traditional lexical matching models are competitive in legal case retrieval tasks. Therefore, we first implement the following lexical matching approach.

\subsubsection{TF-IDF}
TF-IDF~\cite{ramos2003using} is a classical lexical matching model, which is the combination of term frequency (TF) and inverse document frequency (IDF). Their equations are shown as follows:
        \begin{equation}
	TF(t_{i,j}) = \dfrac{n_{i,j}}{\sum_{k}n_{k,j}}
	\label{eq:TF calculation}
        \end{equation}
        
        \begin{equation}
	IDF(t_i) = \log \dfrac{|D|}{|D_i+1|}
	\label{eq:IDF calculation}
        \end{equation}

        \begin{equation}
	TF-IDF = TF \times IDF
	\label{eq:TFIDF calculation}
        \end{equation}
where $D$ is the total number of documents in the corpus and $D_i$ represents the number of documents containing the word $t_i$. $n_{i,j}$ denotes the number of words $t_i$ in the document $d_j$.

\subsubsection{BM25}
BM25~\cite{robertson2009probabilistic} is a probabilistic relevance model based on bag-of-words. Given a query $q$ and a document $d$, the formula of BM25 is shown as follows:

        \begin{equation}
	BM25(d, q) = \sum_{i = 1}^M \dfrac{IDF(t_i) \cdot TF(t_i, d)       \cdot (k_1+1)}{TF(t_i, d) + k_1 \cdot \left(1-b+b \cdot           \dfrac{len(d)}{avgdl}\right)}
	\label{eq:BM25 calculation}
        \end{equation}        
        where $k_1$, $b$ are free hyperparameters, $TF$ represents term frequency and $IDF$ represents inverse document frequency. $avgdl$ is the average length of all documents.

\subsubsection{QLD}
QLD~\cite{zhai2008statistical} is another efficient probabilistic statistical model which calculates relevance scores by considering the probability of query generation. Given a query $q$ and a document $d$, the score of QLD is calculated as follows:

   \begin{equation}
    	\log p(q|d) = \sum_{i: c(q_i; d)>0} \log \dfrac{p_s(q_i|d)}{\alpha_d p(q_i|\mathcal{C})} + n \log \alpha_d +\sum_i \log p(q_i|\mathcal{C})
    	\label{eq:language model calculation}
    \end{equation}

The details can be referred to Zhai et al.'s work\cite{zhai2008statistical}.

\subsection{SAILER}
As mentioned above, legal cases usually contain three parts: Fact, Reasoning, and Decision. Figure \ref{example} illustrates an example of the legal case structure. Key information in the Facts will be carefully analyzed in the Reasoning and influence the final decision. Furthermore, the Reasoning and Decision are written based on the extensive domain knowledge of the judges. Incorporating the rich knowledge inherent in the structure into language models is essential for understanding legal cases.

To achieve the above goals, we propose SAILER~\cite{li2023sailer}, which is shown in Figure \ref{model}. More specifically, SAILER consists of a deep encoder and two shallow decoders. The Fact part is fed to the deep encoder to form a dense vector $h_f$. Then, $h_f$ is concatenated with the positively masked Reasoning and Decision, respectively, which is fed to the shallow decoder. Since the shallow decoder with limited power, $h_f$ is forced to pay more attention to the useful information in the Fact which is relevant to the Reasoning and Decision sections.

To construct the pre-training corpus, we collect 50w legal cases from the U.S. federal
and state courts~\footnote{\url{https://case.law/}}. Then, we extract the corresponding section with regular matching. During the pre-training phase, we optimize the model with the following loss function:
\begin{equation}\label{eqn-6} 
  L_{Total} = L_{MLM} + L_{REA} + L_{DEC}
\end{equation}

\begin{equation}\label{eqn-2} 
  L_{MLM} = -\sum_{x^{'}\in m(F)}\log p(x^{'} | F \backslash m(F))
\end{equation}

\begin{equation}\label{eqn-3} 
  L_{REA} = -\sum_{x^{'}\in m(R)}\log p(x^{'} | [h_{F},R \backslash m(R)])
\end{equation}

\begin{equation}\label{eqn-4} 
  L_{DEC} = -\sum_{x^{'}\in m(D)}\log p(x^{'} | [h_{F},D \backslash m(D))])
\end{equation}
where $F$, $R$, $D$ denote Fact, Reasoning and Decision section respectively. $m(F), m(R), m(D)$ are the masked token of the corresponding section. Only a small percentage of the token (0\%-30\%) in the Fact section is masked since most of the information has to be preserved. The Reasoning and Decision sections have an aggressive masking rate (30\%-60\%) for a better vector representation.

After pre-training, we employ contrastive learning loss to fine-tune. More specifically, given a query case $q$, let $d^{+}$ and $d^{-}$ be relevant and negative cases, the loss function $L$ is formulated as follows:

\begin{equation}\label{eqn-1} 
  L(q,d^+,d^-_{1},...,d^-_{n}) =
-\log_{}{    \frac{exp(s(q,d^+))}{exp(s(q,d^+))+\sum_{j=1}^nexp(s(q,d^-_j))}}
\end{equation}

For each query, we take the irrelevant cases from the top 100 cases recalled by BM25 as negative examples.

\subsection{Learning to Rank}
Following up on previous work~\cite{yang2022thuir,li2023towards,chen2023thuir}, learning to rank techniques are used to further improve performance. In this paper, we integrate all features into the final score with Lightgbm. Table ~\ref{feauture} shows the details of all the features. We employ NDCG as the ranking optimization objective and select the model that performs best on the validation set for testing.

\subsection{Post-processing}
After getting the ranking scores, we perform the following post-processing strategy:
\subsubsection{Filtering by trial date}
Since query cases can only cite cases that are judged before itself, we filter the candidate set according to trial date. Specifically, we extract all the dates in the case, i.e., four digits within a reasonable range. Then, the largest date that appears is regarded as the trial date of the case. This avoids wrong filtering caused by treating other dates as the trial date. If the trial date of the query case is unknown, its candidate set contains all other cases.

\subsubsection{Filtering query cases}
We note that the average number of times that query cases are noticed is 0.056 in the training set. Therefore, after getting the relevant cases for each query, we delete all query cases included in it.

\subsubsection{Dynamic cut-off}
It is noticeable that the number of cases relevant to each query case is variable. Therefore we employ dynamic cut-off to identify the relevant cases for each query.
We define $l$ as the minimum number of noticed cases and $h$ as the maximum number of noticed cases. After that, we take the highest score $S$ as the basis, and only cases with scores greater than $p \times S$ are returned. Grid search is performed on the validation set to determine the optimal value of $p, l, h$.

\section{Experiment}
We conduct experiments to verify the effectiveness of our proposed method. Specifically, this section investigates the following research questions:
\begin{itemize}[leftmargin=*]
    \item \textbf{RQ1}: What are the advantages of SAILER over the previous pre-trained and lexical matching models?
    \item \textbf{RQ2}: How do different post-processing strategies affect final performance?
\end{itemize}

\subsection{Implementation Details}
For traditional lexical matching models, we implement them with the pyserini toolkit~\footnote{\url{https://github.com/castorini/pyserini}}. We notice that BM25 does not perform well with the default parameters, so we set $k_1=3.0$ and $b=1.0$.  

For pre-training, the masking rate of the encoder is 0.15, and the masking rate of decoders is 0.45. We pre-train up to 10 epochs using AdamW~\cite{loshchilov2018fixing} optimizer, with a learning rate of 1e-5, batch size of 72, and linear schedule with warmup ratio of 0.1. In the fine-tuning process, the ratio of positive to negative samples is 1:15. We fine-tune up to 20 epochs using the AdamW~\cite{loshchilov2018fixing} optimizer, with a learning rate of 5e-6, batch size of 4, and linear schedule with warmup ratio 0.1. All the experiments in this work are conducted on 8 NVIDIA Tesla A100 GPUs.

For learning to rank, we set the learning rate to 0.01, the number of leaves to 20, and the early stopping step to 100. The boosting\_type is ``gbdt" and the objective is ``lambdarank". During post-processing, $l/h$ are eventually 4/6 respectively, and $p$ is set to 0.84.

\begin{table}[t]
\caption{Performance of single model on COLIEE 2023 validation set.``-" represents the unlimited length.}
\begin{tabular}{lcccc}
\hline
model             & \multicolumn{1}{l}{max\_length} & \multicolumn{1}{l}{P@5} & \multicolumn{1}{l}{R@5} & \multicolumn{1}{l}{F1 score} \\ \hline
BM25(k\_1=3,b=1)              & 512                             & 0.0963                  & 0.1067                  & 0.1012                       \\
QLD               & 512                             & 0.0983                  & 0.1091                  & 0.1035                       \\
BERT        & 512                             & 0.0770                  & 0.0854                  & 0.0809                       \\
RoBERTa     & 512                             & 0.0994                  & 0.1103                  & 0.1046                       \\
LEGAL-BERT        & 512                             & 0.0845                  & 0.0937                  & 0.0888                       \\
SAILER      & 512                             & 0.1315                  & 0.1459                  & 0.1385                       \\

TF-IDF            & -                               & 0.0898                  & 0.1504                  & 0.1142                       \\
BM25(k\_1=3,b=1)         & -                               & 0.1465                  & 0.1625                  & 0.1541                       \\
QLD               & -                               & 0.1411                  & 0.1565                  & 0.1484                       \\ \hline
\end{tabular}
\label{valid}
\end{table}

\begin{table}[]
\caption{Ensemble with different post-processing strategies}
\begin{tabular}{lccc}
\hline
model                    & \multicolumn{1}{l}{P@5} & \multicolumn{1}{l}{R@5} & \multicolumn{1}{l}{F1 score} \\ \hline
Ensemble                 & 0.1863                  & 0.2032                  & 0.1944                       \\
+Filtering by trial date & 0.2070                  & 0.2290                  & 0.2175                       \\
+Filtering query cases   & 0.2092                  & 0.2314                  & 0.2197                       \\
+Dynamic cut-off         & 0.2177                  & 0.2385                  & 0.2276                       \\ \hline
\end{tabular}
\label{ensemble}
\end{table}

\subsection{Experiment Result}
To answer \textbf{RQ1}, we compare the performance of different single models and analyze the strengths and weaknesses of pre-trained language models. Table \ref{valid} shows the performance comparison of the different methods. We can get the following observations:

\begin{itemize}[leftmargin=*]
    \item When the input lengths of the models are the same, the performance of RoBERTa~\cite{liu2019roberta} is approximate to that of BM25 and QLD. Since there are no pre-training tasks designed for dense retrieval, LEGAL-BERT~\cite{chalkidis2020legal} does not achieve competitive performance.
    \item Benefiting from the expert knowledge inherent in the structure of legal cases, SAILER outperforms traditional lexical matching models and pre-trained language models under the same conditions.
    
    \item  However, the performance of BM25 and QLD is further improved when the input length is not limited. The traditional lexical matching model is still competitive under long-text legal cases. The input length limits the further understanding of the legal instrument by language models. In the future, we will continue to explore the performance of language models based on Longformer for legal case retrieval.
\end{itemize}

To answer question \textbf{RQ2}, we employ different post-processing strategies on the score of ensemble. From the experimental results in Table \ref{ensemble}, we can obtain the following observations:

\begin{itemize}[leftmargin=*]
    \item Compared with the effectiveness of single models, learning to rank incorporates multiple features and achieves further performance improvements.
    \item All three post-processing strategies facilitate performance improvement. Narrowing the candidate set for each query via the strategy of filtering by trial date achieves the best boosting effect.
\end{itemize}

The final top-5 results of COLIEE 2023 Task 1 are illustrated in Table \ref{test}. Our run2 has the best performance and is significantly better than other runs. Run 3 and Run 1 are other processing methods with different parameters. Finally, the THUIR team wins the championship.

\begin{table}[]
\caption{Final top-5 of COLIEE 2023 Task 1 on the test set.}
\begin{tabular}{ccccc}
\hline
\textbf{Team} & \textbf{Submission} & \textbf{Precision} & \textbf{Recall} & \textbf{F1}  \\ \hline
THUIR         & thuirrun2                & 0.2379             & 0.4063          & 0.3001                                      \\
THUIR         & thuirrun3                & 0.2173             & 0.4389          & 0.2907                                      \\
IITDLI        & iitdli\_task1\_run3      & 0.2447             & 0.3481          & 0.2874                                      \\
THUIR         & thuirrun1                & 0.2186             & 0.3782          & 0.2771                                     \\
NOWJ          & nowj.d-ensemble          & 0.2263             & 0.3527          & 0.2757                                     \\ \hline
\end{tabular}
\label{test}
\end{table}
\section{Conclusion}
This paper presents THUIR Team’s approaches to the legal case retrieval task in the COLIEE 2023 competition. Due to the limited training data, we employ a legal-oriented pre-training model to improve performance. Furthermore, diverse pre-processing and post-processing approaches are presented. Also, we utilize learning to rank to merge the different features into the final score. Finally, we win first place in this competition. In the future, we will explore more pre-training objectives suitable for legal case retrieval.
    
\bibliographystyle{ACM-Reference-Format}
\bibliography{sample-base.bib}
\end{document}